\documentclass{optica-article}

\journal{opticajournal} 

\articletype{Research Article}

\usepackage{breqn}
\usepackage{lineno}

\newcommand{\figref}[1]{Fig.~\ref{fig:#1}}
\newcommand{\tabref}[1]{Table~\ref{tab:#1}}
\newcommand{\eqnref}[1]{Eq.~\eqref{eq:#1}}

\newcommand{\abs}[1]{\left|#1\right|}

\newcommand{\diff}{{\rm d}}

\newcommand{\um}{\mu{\rm m}}

\makeatletter
\renewcommand{\paragraph}{%
  \@startsection{paragraph}{4}%
  {\z@}{0.60ex \@plus 1ex \@minus .15ex}{-1em}%
  {\normalfont\normalsize\bfseries}%
}
\makeatother
\begin{document}

\title{Metalens Enhanced Ray Optics: An End-to-End Wave-Ray Co-Optimization Framework}

\author{Ziwei Zhu,\authormark{2, *} Zhaocheng Liu,\authormark{1} and Changxi Zheng\authormark{2}}

\address{\authormark{1}Meta Reality Labs, 9845 Willows Rd NE, Redmond, Washington 98052, USA\\
\authormark{2}Department of Computer Science, Columbia University, New York, New York 10027, USA}

\email{\authormark{*}zz2556@columbia.edu} 


\begin{abstract*} 
    We present a fully differentiable framework for seamlessly integrating wave optical components with geometrical lenses, offering an approach to enhance the performance of large-scale end-to-end optical systems. In this study, we focus on the integration of a metalens, a geometrical lens, and image data. Through the use of gradient-based optimization techniques, we demonstrate the design of nonparaxial imaging systems and the correction of aberrations inherent in geometrical optics. Our framework enables efficient and effective optimization of the entire optical system, leading to improved overall performance.
\end{abstract*}

\section{Introduction}
The end-to-end method~\cite{sitzmann2018end, lin2021end}, as a novel optical design paradigm, has emerged recent years because of the improving of computer hardwares and the need of new imaging capabilities.~\cite{nayar2006computational} This method offers advantages for user customization and more flexible and intuitive figure-of-merits. Compared to the conventional strategy~\textemdash~designing each optical component and combining them together, it greatly improves the overall performance in a system level. It can also refine the system designed through the conventional strategy.

An optical system typically consists of lots of components. The design of these components is guided by wave optics~\cite{miyamoto1961phase, pan2022dielectric} or geometrical optics~\cite{cakmakci2021holographic}. The behavior of wave optical components can be solved using first-principle Maxwell's equations, which is accurate and opens the possibility of exotic optical phenomena such as negative refractive index~\cite{smith2004metamaterials} and ultra broadband behaviors~\cite{cheng2022genetic}. However, direct simulations based on those accurate principles are often costly and cannot scale well.

Contrast to wave optics, geometrical ray tracing is more applicable for larger system. It is also easy to model long propagation distance and large incident angle by ray tracing. An extreme example is the rendering techniques used in game engine such as Unreal~\cite{hart2021practical}.

Recent years, lots of progress has been made in the differentiable simulation of wave optics~\cite{hughes2019forward, zhu2020differentiable} and geometrical optics~\cite{zhang2019differential, teh2022adjoint}. It is desired to combine the exotic properties of wave optics and the scalability of geometrical optics together. With that, it is possible to model nonparaxial imaging system with large incident angle, or correct aberrations in geometrical optics. 

However, this combination is not straight-forward. The description of light in the above optical systems is fundamentally different: In a wave optical system, the light is described by a field.~\cite{sullivan2013electromagnetic}, whereas in geometric optics, the light is described by discrete rays. The conversion itself experiences information loss, if no high-order model of rays is used. Also, this conversion can be hard to differentiate, if not impossible, because of the discretization of rays.

\begin{figure*}
\includegraphics[width=0.9\textwidth]{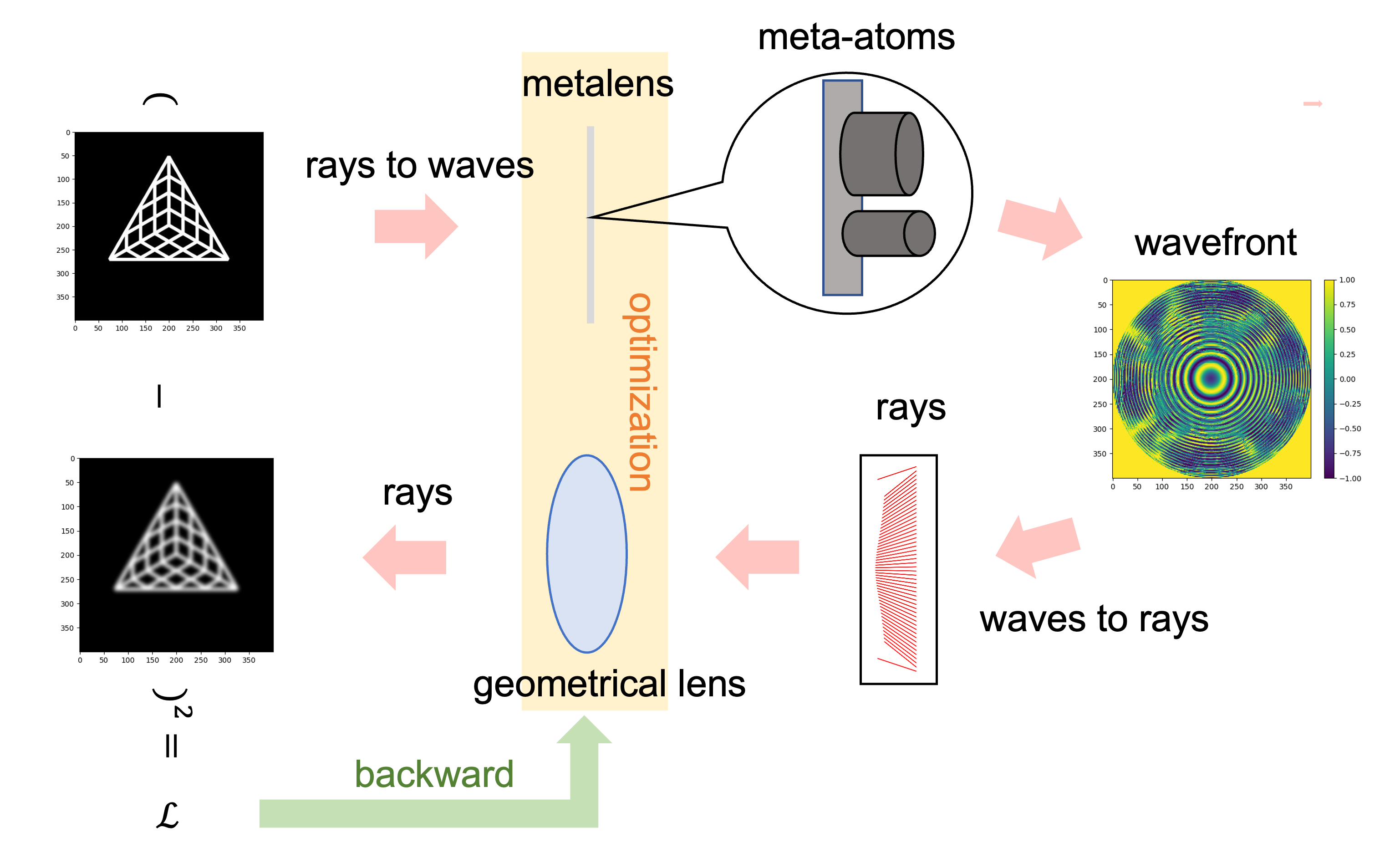}
\caption{The end-to-end wave-ray co-optimization framework. The inputs consist of plane waves from multiple directions, which are used to construct the input wavefront. The wavefront then passes through a metalens and is converted back to rays, which are directed through a geometrical lens to simulate the point spread functions. From these functions, we can simulate the output image. The optimization process involves jointly optimizing the metalens and geometrical lens to minimize the difference between the input and output images.}\label{fig:end_to_end_framework}
\end{figure*}

We aim to develop differentiable wave-ray conversion algorithms that can be easily integrated with state-of-the-art wave simulation and ray tracing softwares. Based on that, we introduce an optimization framework for a hybrid system consists of both wave optical components and geometrical lenses. The output image is also reconstructed by point spread functions,  which can be easily combined with image reconstruction algorithms and deep neural networks. The proposed framework is outlined in \figref{end_to_end_framework}.

As illustrated in \figref{end_to_end_framework}, the process begins by sourcing rays from the input image, which then accumulate to form the incident wavefront. This wavefront subsequently passes through a wave optical component, following which the rays are extracted and directed through a geometrical lens. This setup enables simulation of the output image for any focal point and assessment of how the output image changes as parameters of the wave optical component or the geometrical lens are varied. By minimizing the difference between the output image and the input image, an end-to-end optimization pipeline can be constructed. As examples, we show how to use this system for nonparaxial optical optimization and minimizing color aberrations.

\section{Optimization Framework}
\subsection{Mathematical Model}
To begin, we assume that the light is described by a perfect plane wave when observed from a far-field perspective. Let us assume that the plane wave has a uniform wave vector $\vec{k}_{\text{in}}$. Consequently, the input wavefront at a position $\vec{r}$ can be represented by the equation:
\begin{equation}
\phi_{\text{in}}(\vec{r}) = \exp{\left(j\vec{k}_{\text{in}}\cdot\vec{r}\right)}.
\end{equation}
It is worth noting that our framework is also applicable to point sources. In this scenario, the input wavefront would be $\phi_{\text{in}}(\vec{r}) = \exp{\left(jk_0\Vert \vec{r}_{\text{in}} - \vec{r}\Vert \right)}$, where $\vec{r}_{\text{in}}$ is the position of the point source and $k_0$ is the wavenumber in vacuum.

As the wavefront passes through a flat metalens, the modulation of the wavefront can be described as a combination of the amplitude $A$ and phase $S$ modulations:
\begin{equation}
\phi_{\text{mod}} = A(\vec{r})\exp{\left(jS(\vec{r})\right)}.
\end{equation}
Since the thickness of the metalens is negligible compared to the scale of the entire system, it can be ignored. The modulation of amplitude and phase caused by the metalens, along with its gradients with respect to the geometries of meta-atoms, are modeled by a differentiable neural network, such as SIREN.~\cite{sitzmann2020implicit} The main reason for the surrogation is that first-principle solver is oftentimes much more costly compared to other parts of our system.

The output wavefront can be computed by multiplying the input wavefront and the modulation wavefront:
\begin{equation}
\phi_{\text{out}} = \phi_{\text{mod}}\phi_{\text{in}}.
\end{equation}
Our objective is to extract rays from the output wavefront $\phi_{\text{out}}$.
\subsection{Differentiable Conversion from Waves to Rays}
It is necessary to assume the light can be represented by a linear combination of plane waves for the conversion. In other words, the light is in pure state.~\cite{isham2001lectures} For light in mixed state, it is impossible to extract plane waves only from wavefront in frequency domain.

There are muliple ways to convert waves to rays. We proposed two methods that can be combined with differentiable ray tracing, which means, the output must be some quantities with definite directions and starting positions. We summarized the advantages and limitation of each method in \tabref{summary_methods}.
\begin{table}[!htp]
    \centering
\begin{tabular}{cccc}
    \hline
    method    &  \#rays    & accuracy & single plane wave\\
    \hline
    windowed Fourier transform &  $n$ & + & no\\
    phase gradient method & $1$ & ++ & yes\\
    \hline
\end{tabular}
\caption{The summary of the two wave to ray conversion methods. Windowed Fourier transform can output multiple rays, at the cost of the blurriness caused by the finite size of the window. Phase gradient method is free from blurriness but can only output one ray direction, this is the only output ray direction given one input direction and uniform amplitude. }\label{tab:summary_methods}
\end{table}

As is shown in \tabref{summary_methods}, windowed Fourier transform does not assume the field is a plane wave locally, instead, it linearly decomposes light into several plane waves. However, due to the uncertainty principle, the size of the window in the windowed Fourier transform is inversely proportional to the uncertainty of the output ray direction. If we use phase gradient method, it will be free from this uncertainty caused by the finite window size.
But it assumes the phase of the wavefront is well-defined. In other words, for each point at the wavefront, we assume it can be locally viewed as a plane wave.
\subsection{Differentiable Ray Tracing}
Once rays are extracted from the wavefront, the rays can be passed into the geometrical lens for ray tracing.~\cite{shirley2008realistic} This is at the cost of overlooking the diffraction, which is sufficient for visible light over long distance at the scale of several milimeters, given a large enough aperture.~\cite{airy1835diffraction} 

An optical component often has smooth boundary, which means its bidirectional reflectance distribution function~\cite{nicodemus1965directional} is a delta function, or a summation of several delta function. The tracing process can be modeled sequentially by a unified method \textbf{trace} for different objects. Each time, the \textbf{trace} function takes the input positions $\vec{r}$s and wavevectors $\vec{k}$s, and output the the hitting positions and the refracted wavevectors. 

For efficiency concern, both input and output are treated as large tensors consisting of the information from all the rays. We neglect reflection. Once the total reflection happens, we set the intensity of the correponding ray to be zero, and its position can be arbitrary. By overlooking the branching and pruning the rays far away from the optical axis, we are able to vectorize the process and take advantage of GPU computation of PyTorch to compute more than $10^4$ rays. 

\subsection{Image Simulation with Differentiable Point Spread Function}
Once a ray hits the image plane at a sensor pixel, the sensor will record a signal. Because we adopt the ray tracing model, diffraction pattern and interference is overlooked. However, the postions of pixels are fixed in space, which is not differentiable inherently. To develop a differentiable point spread function for end-to-end image optimization, we first write the point spread function as a summation of delta functions:
\begin{equation}
    \text{PSF}(\vec{r}) = \frac{1}{B}\sum_{k}\left|A_k\right|^2\delta(\vec{r} - \vec{r}_k)
\end{equation}
where $B$ is a normalization term to ensure the PSF function sums to $1$. 

At the first glance the delta functions in PSF are not differentiable inherently. In practice, we can either fit the PSF using Gaussian function~\cite{hale2021end} or relax the delta function as a function with small variance. However, the relaxation introduces non-physical biases. Also, if the variance is too small, the gradients approach zero far away from the hit points and infinity near the hit points. 

We develop the differentiable PSF by considering output image $G(\vec{r})$ as a convolution of the input image $F(\vec{r})$ with the spatially variant PSF
\begin{dmath}
    G(\vec{r}) = F(\vec{r}) * \text{PSF}(\vec{r}) = \frac{1}{B} \int\int \sum_{k}\left|A_k\right|^2\delta(\vec{r}' - \vec{r}_k)F(\vec{r} - \vec{r}')\diff s(\vec{r}')
    = \frac{1}{B}\sum_{k}\left|A_k\right|^2 F(\vec{r} - \vec{r}_k)
\end{dmath}

Although the input image is always discretized, for any position not aligned with the pixel grid, the color value can be found by an interpolation of the neighbor grid cells. This inspires us to precompute a PSF kernel. If the image size is $N_y\times N_x$, the size of the PSF kernel should be $(2N_y-1)\times(2N_x-1)$ to cover the whole image, even when convoluting at the corner of the image. If the ray hits out of the kernel, it will never contribute to the signal of the sensor.

Now we describe the geometrical PSF as follows: we initialize the PSF as a zero matrix. Each time when the ray hits $\vec{r}_k$, assuming $\vec{r}_k$ is between four pixels $(i, j)$, $(i, j+1)$, $(i+1, j)$, and $(i+1, j+1)$ on the PSF kernel, and we have
\begin{eqnarray}
    \alpha_k &=& \frac{y_k - y_i}{y_{i+1} - y_i}\\
    \beta_k &=& \frac{x_k - x_j}{x_{j+1} - x_j},
\end{eqnarray}
we can add elements to the PSF kernel as follows:
\begin{eqnarray}
    P(i, j) &\mathrel{+}=& \sum_k\abs{A_k}^2(1-\alpha_k)(1-\beta_k)\\
    P(i, j+1) &\mathrel{+}=& \sum_k\abs{A_k}^2(1-\alpha_k)\beta_k\\
    P(i+1, j) &\mathrel{+}=& \sum_k\abs{A_k}^2\alpha_k(1-\beta_k)\\
    P(i+1, j+1) &\mathrel{+}=& \sum_k\abs{A_k}^2\alpha_k\beta_k
\end{eqnarray}

The resulting PSF is differentiable thanks to the assumption that the image is smooth and can be interpolated. It is direct to get the gradients of $\alpha_k$ and $\beta_k$ with respect to $\vec{r}_k$. The final convolution should have no padding, so the output will be of the same size as the input image. If we consider the spatial variance of the PSF, we can precompute a sample of PSFs (like $9\times 9$), the output image will be the convolution of the interpolated PSFs and the input image. 

\subsection{Optimization and Practical Consideration}

The image reconstruction loss through the optical system can be formulated as
\begin{equation}
    \mathcal{L} = \frac{1}{S}\int\int\Vert F(\vec{r}) - G(\vec{r})\Vert^2_2\diff s
\end{equation}
where $S$ is the area of the image. This is working for both paraxial PSF and nonparaxial PSFs. For multiple channels like RGB, we sum the loss functions of different colors together. 

However, directly optimizing the image can get stuck in a local minimum. We find it is better to get the initial guess by optimizing the paraxial property first. In order to do this, we optimize the hit point positions $\vec{r}_k$ on the image plane with a perpendicular incident plane wave first. The position-based loss can be formulated as 
\begin{equation}\label{eq:position_based_loss}
    \mathcal{L} = \sum_k \Vert \vec{r}_k -\vec{r}_0\Vert^2_2 
\end{equation}
where $\vec{r}_0 = (0, 0)$ is the position of the focal point.

\section{Windowed Fourier Transform}
A natural idea to convert from waves to rays is to decompose the wavefront into multiple plane waves. For different plane waves, we sample infinite number of rays along the corresponding direction at different positions. This is impossible in practice. Therefore we need to choose a window for the Fourier transform.

Windowed Fourier transform is widely used in audio processing to extract different pitches of sound at a certain time $t$.~\cite{moorer2017note}, which is also called short-time Fourier transform. For light waves, we need to consider the whole output plane, slided by a 2D window. At any position $\vec{r}$, the ray intensity pointing to an arbitrary wavevector $\vec{k}$ is given by
\begin{equation}
 {\abs{A(\vec{r}, \vec{k})}}^2 = \left[\int_{\Omega} \phi_{\text{out}}(\vec{r}')\exp{\left(-j\vec{k}\cdot(\vec{r}' - \vec{r})\right)}\diff s(\vec{r}')\right]^2
\end{equation} 
where $\Omega$ is the 2D window. From the equation, we know the result is blurred by the Fourier transform of the aperture $\Omega$~\textemdash~the smaller the aperture is, the result is blurrier. However, increasing the window size will delocalize the position of the ray~\textemdash~increase the width of the light beam. In summary, the window size is a hyperparameter, and we need to carefully tune the window size to ensure transformed rays are close to actual ray propagation. 

\begin{figure}
    \centering
    \includegraphics[width=0.9\textwidth]{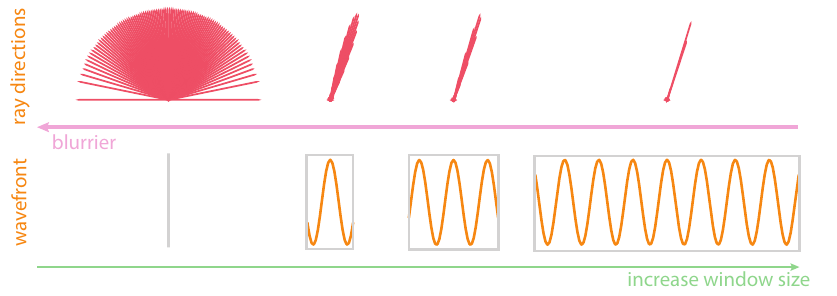}
    \caption{Visualization of the blurriness caused by finite window size. It is shown that by increasing the window size, the blurriness will be minimized. }\label{fig:illustrate_uncertainty}
\end{figure}

To illustrate the blurriness, we adopt a simple example, a 1D wavefront generated by a uniform wavevector $\vec{k}$. We extract different amplitudes along different directions using windowed Fourier transform. We find that, the smaller the window size is, the directions are blurrier. However, the output should only have one direction. The visualization is shown in \figref{illustrate_uncertainty}.

The gradient can be backpropagated via a similar formula

\begin{equation}
    \frac{\partial \mathcal{L}}{\partial \phi^*_{\text{out}}(\vec{r})} = \int_{\Omega} \left[ \frac{\partial \mathcal{L}}{\partial A(\vec{r}', \vec{k}')}\exp{\left(j\vec{k}'(\vec{r}' - \vec{r})\right)}\right]^*\diff s(\vec{r}')
\end{equation}
where $\vec{k}'$ is calculated using the target point and $\vec{r}'$. Notice that for backpropagation, we utilize Wirtinger derivative~\cite{hunger2007introduction}, which is compatible with PyTorch. 

\begin{figure}[!htp]
    \centering
    \includegraphics[width=0.9\textwidth]{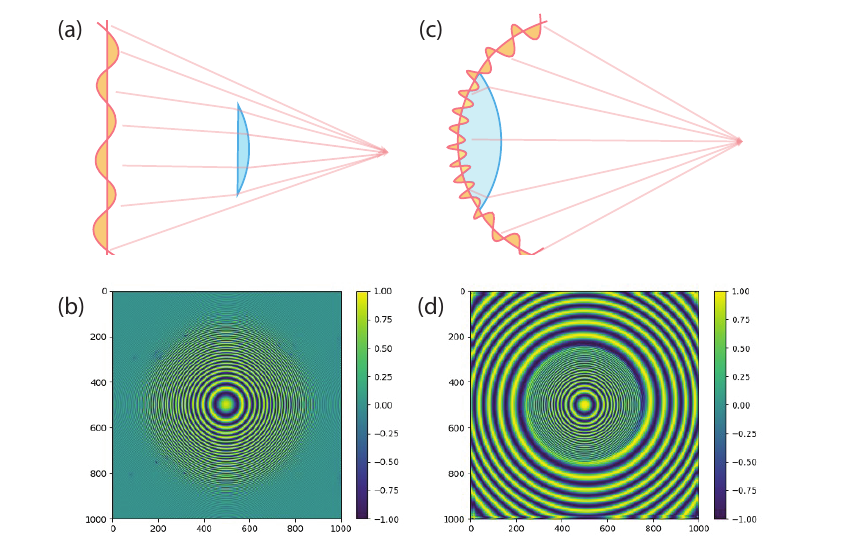}
    \caption{The co-optimization of metalens and geometric lens. (a) The metalens is seperated from the convex lens by $2 {\rm mm}$. The curvature radius of the convex lens is fixed $3 {\rm mm}$ and the optimized wavefront $\phi_{\text{mod}}$ is shown in (b). (c) The metalens is coated at one side of the convex lens, and the curvature radius of both side of the convex lens is optimized from $3 {\rm mm}$ to $3 {\rm mm}$. The optimized wavefront on the metalens is shown in (d). For simplicity we only optimize the center PSF.}\label{fig:metalens_and_geometric_stft}
\end{figure}

The outputs of the windowed Fourier transform are the amplitudes of rays pointing to arbitrary directions $\vec{k}$. As an approximation, it is possible to extract the ray directions with top-k amplitudes. And the differentiation of the top-k problem is solvable by solving an optimal transport problem.~\cite{xie2020differentiable} This method, however, is very slow even implemented on GPUs. Another way is to decompose the optimization into two steps. In the first optimization step, we solve the directions of the output rays that will optimize the imaging quality. Then in the second optimization step we optimize the amplitude and phase of the metalens to maximize the amplitude of the rays pointing to the target directions. We demonstrate the optimization using two setups: In the first setup, the phase modulation is coated at one side of the convex lens; in the second setup, the phase modulation is seperated from the convex lens. We assume the incident waves only come from the direction perpendicular to the flat metalens, so the incident wavefront is uniform. The setup and resulting wavefront is shown in \figref{metalens_and_geometric_stft}.

As is shown in \figref{metalens_and_geometric_stft}, our algorithm are able to optimize the convex lens together with the metalens. For simplicity, we only optimize the plane waves that is perpendicular to the wavefront without considering different plane waves from different angles. We can observe that, inside and outside of the aperture by the geometric lens, the wavefront $\phi_{\text{mod}}$ is different. This is because if the rays coming from the metalens hit nothing, it will go straightly. However, if the rays hit the convex lens, their directions change.

\section{Phase Gradient Method}
Consider the change of amplitude will split the rays into multiple directions, oftentimes we only need to optimize the phase modulation for a better result. We assume the wave is described by a scalar field:
\begin{equation}
    \phi_{\text{out}} = \exp{\left(jS_{\text{out}}(\vec{r})\right)}
\end{equation}

If the wave optics component acts like a phase modulator, it will only change $S_{\text{out}}$. The phase gradient will deflect the direction of the output rays, according to the generalized refraction law.~\cite{yu2011light} If we assume the output wave is plane wave, we propose the \textbf{equivalent wavevector extractor}:
\begin{equation}\label{eq:wavevector_extractor}
    \vec{k} = \frac{\partial S_{\text{out}}}{\partial \vec{r}}
\end{equation}

Note that here we assume that at each position, only one $\vec{k}$ is possible. This is working for coherent plane waves without amplitude modulation, or as an approximation of the average light propagation direction for incoherent light. 

To implement the equivalent wavevector extractor, assume the phase is discretized by a uniform grid with stepsize $h$, the three components of the wavevector are
\begin{eqnarray}
    k_x(i, j) &=& 0.5(S(i, j+1) - S(i, j-1))/h\\
    k_y(i, j) &=& 0.5(S(i+1, j) - S(i-1, j))/h\\
    k_z(i, j) &=& \sqrt{k^2_0 - k^2_x(i, j) - k^2_y(i, j)}
\end{eqnarray}
By implementing the steps in PyTorch, we can use Autograd~\cite{paszke2019pytorch} to compute the gradient of $\vec{k}$ with respect to the phase $S$.

\section{Results}
\subsection{Paraxial and Nonparaxial Optical Design}
Firstly, we validate the design capability of our framework by co-optimize a setup with both optical phase modulator and convex lens. We adopt the position-based loss first. 

\begin{figure}
    \centering
    \includegraphics[width=0.7\textwidth]{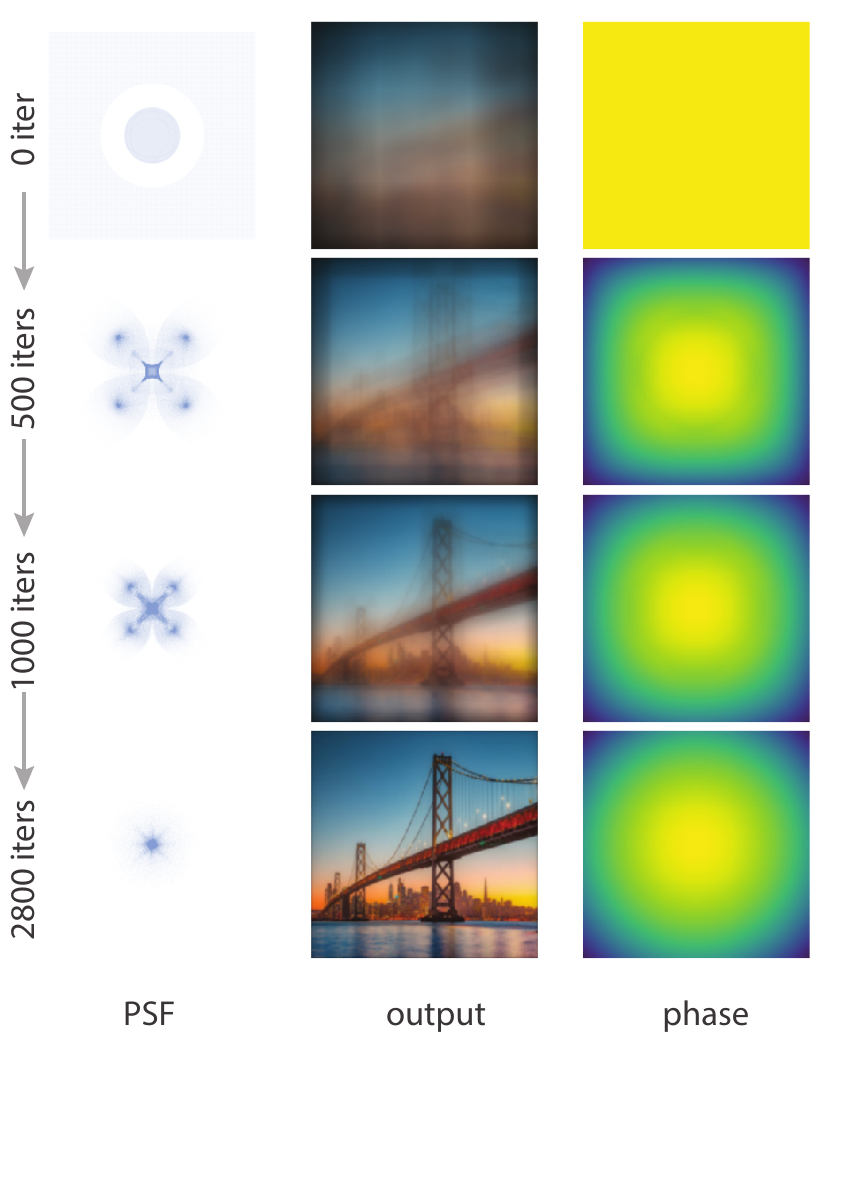}
    \vspace{-15mm}
    \caption{To improve the near axis point spread function, we utilize a position-based loss function, as described in \eqnref{position_based_loss}. We begin by applying a uniform phase modulation, then allowing the output rays to pass through a convex lens with a fixed curvature radius of $3~{\rm mm}$. The left column depicts the progression of the point spread function for the entire system. The middle column showcases the resulting output image after passing through the hybrid optical system. The right column displays the phase modulation during the optimization process.}\label{fig:paraxial_position_based_opt}
\end{figure}
The results are shown in \figref{paraxial_position_based_opt}. The phase lens has a diameter of $2{~\rm mm}$. And we assume the geometrical lens is placed at $2~{\rm mm}$ away from the phase lens, with an aperture of $0.5~{\rm mm}$. The focal point is $5~{\rm mm}$ away from the phase lens. We optimize to let all the rays from the wave plate hit the image plane at the center. Notice that this setup is only for demonstration. For nonparaxial optical design, we need to make sure the aperture of the geometrical lens is large enough, so plane waves from large incident angles will still pass the convex lens. We assume the radius of the convex lens is fixed as $3~{\rm mm}$. 

From~\figref{paraxial_position_based_opt}, we can observe that the initial PSF consists of a small focused circle, surrounded by a ring. This is because the  light perpendicularly incident to the convex lens. The rays entering the aperture will be focused while the rays out of the aperture will not change their directions. We show that by optimizing the whole system, we are able to focus both rays to the focal point. For simulation, we only consider the green light ($520~{\rm nm}$). It can be easily extended to cover other wavelengths by adding the corresponding loss functions. 

\begin{figure}
    \centering
    \includegraphics{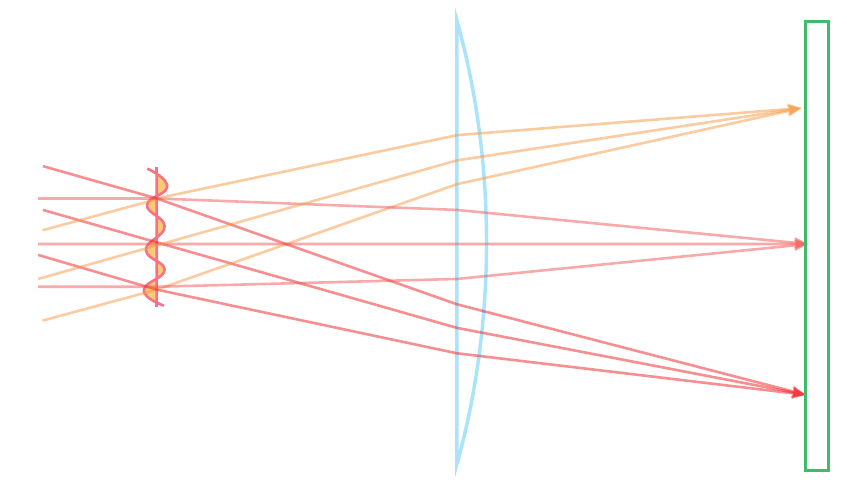}
    \caption{This figure depicts the device setup for the nonparaxial optical design. Plane waves from various directions undergo an optimized phase modulation of $2~{\rm mm}$ and then pass through a large convex lens. The convex lens aperture is maintained at $7.9~{\rm mm}$ to allow plane waves from incident angles as high as $\pm 12^\circ$ to pass through. The phase modulation and convex lens are separated by a distance of $2~{\rm mm}$, and the sensor and convex lens are also separated by a distance of $2~{\rm mm}$. We perform joint optimization of the phase modulation and convex lens to enhance the imaging quality on the sensor from the far field.}\label{fig:nonparaxial_optical_design}
\end{figure}

Next, we consider a more realistic scene with light incident from large angles. For each incident angle, the light is modeled by a plane wave with a fixed wavevector $\vec{k}$. Because our setup consists of two lenses, one phase modulator and one geometrical lens, to ensure the rays from the modulator passing through the second lens, the aperture of the geometrical lens should be large enough. An illuration of the setup is shown in figure~\figref{nonparaxial_optical_design}.

For each plane wave, the phase at any coordinate when incident at the metalens can be collected easily. Assume the $xy$-coordinates of the metalens is $x_{\text{inc}}$ and $y_{\text{inc}}$, and the wavevector of the plane wave is $\vec{k}_{\text{inc}} = (k_x, k_y)$, the incident phase can be formulated as
\begin{equation}
    S_{\text{inc}} = k_x x_{\text{inc}} + k_y y_{\text{inc}}
\end{equation}
If the phase contribution of the optimizable modulator is $S_{\text{mod}}(x, y)$, the output phase can be written as
\begin{equation}
    S_{\text{out}} = S_{\text{inc}} + S_{\text{mod}}
\end{equation}

This output phase can be put into \eqnref{wavevector_extractor} to get the output wavevector $\vec{k}$. After the ray tracing, each plane wave instance will hit a set of points at the imaging plane. If we optimize for an image with pixel size $p = 13~\um$, the sensor size is assumed to be $pN_y\times pN_x$. In practice, we sample the centers of PSFs uniformly within the sensor. 

\begin{figure*}[!htp]
    \centering
    \includegraphics[width=\textwidth]{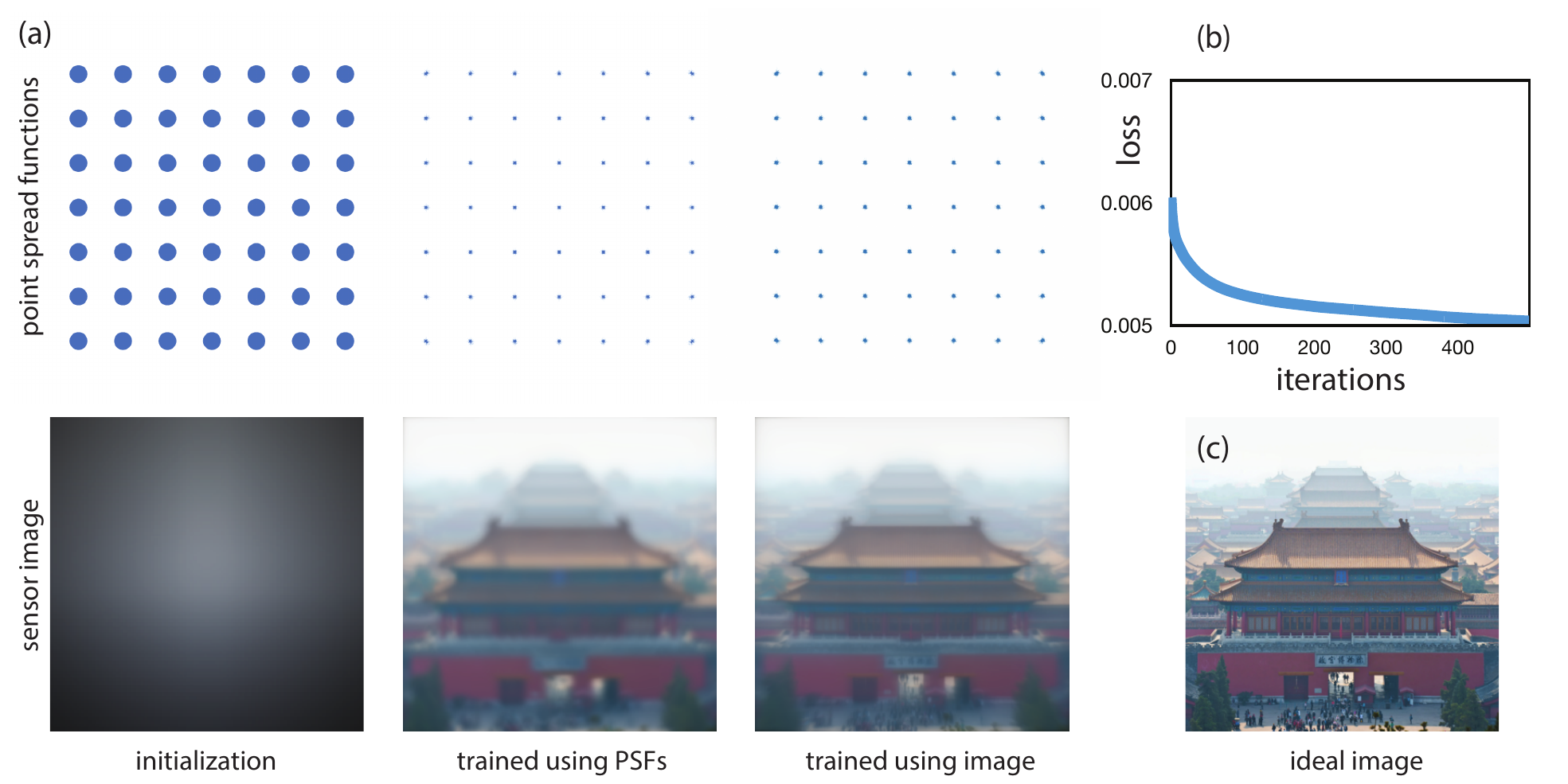}
    \caption{(a) To optimize the point spread functions and the resulting image, we first set the lens radius to $400~{\rm mm}$ and initialize the phase modulation to be uniform. Then, we train both the geometrical lens and the wave optical component by minimizing the deviation of the points in the simulated point spread functions from their corresponding target centers for the plane waves. Finally, we fine-tune the results by minimizing the difference between the input image and the output image. (b) The loss - iterations curve of the fine tuning process. (c) The ideal output image, which is the input of our optical system. }\label{fig:off_axis_results}
\end{figure*}

We optimize the curvature radius of the geometrical lens and the phase modulation together. Because directly optimizing on image can easily get stuck in local minima, in practice, we optimize to minimize the spreading in the geometrical point spread functions (spot diagrams) firstly. However, the optimization directly based on this loss is not sufficient because different point spread functions may have different contributions to the final imaging quality, and it is hard to determine the weights between different point spread functions. Also, it is noticed that different regions in the image may have difference importance. For example, the region with more details may need a better point spread functions. Based on this, we use the result as a good initial guess to optimize the image. The optimized point spread functions and the resulting image are shown in \figref{off_axis_results}.

\begin{figure*}[!htp]
    \hspace{7mm}
    \includegraphics[width=0.95\textwidth]{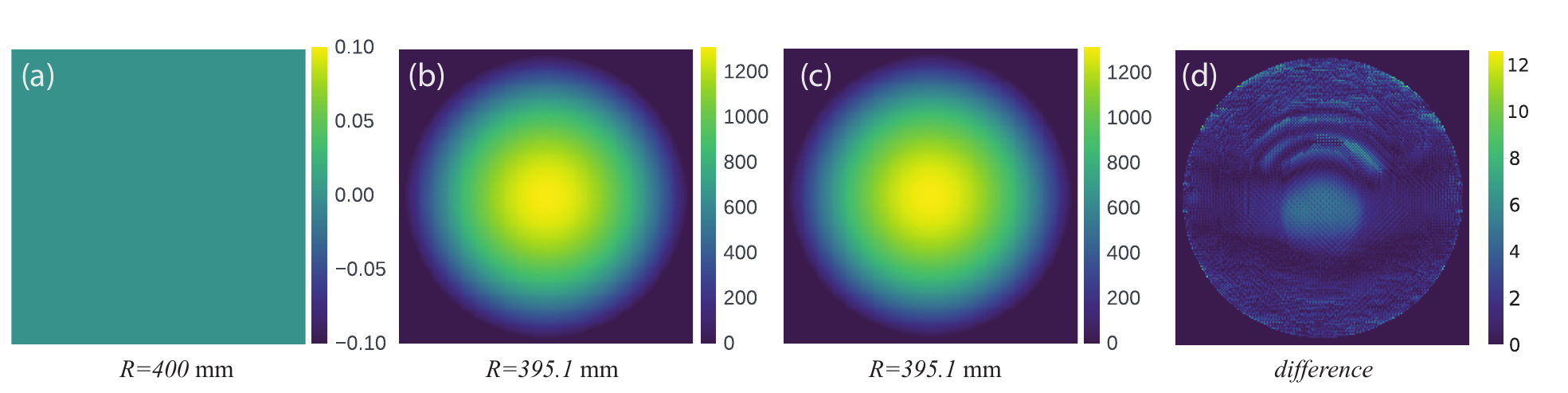}
    \caption{The optimization progress of phase modulation and curvature radius for the geometrical lens. (a) The initial state with a uniform phase modulation and a lens radius of $400~{\rm mm}$. (b) The optimized phase modulation and curvature radius obtained by minimizing the deviation of the geometrical point spread functions from their target centers. (c) The further refined phase modulation and lens radius obtained by minimizing the difference between the input and output images. (d) The difference between(b) and (c).}\label{fig:phase_opt_off_axis}
    \vspace{-5mm}
\end{figure*}

To further analyse the results, we also visualize the phase and lens radius during the optimization, which is shown in \figref{phase_opt_off_axis}. 

\subsection{Enhanced Geometrical Optics for Virtual Reality}
Nowadays, most virtual reality (VR) systems rely on geometrical optics for guidance~\cite{cakmakci2021holographic}. However, the use of simple lenses in these systems often leads to errors such as spherical and color aberrations due to the limitations of traditional optics and manufacturing techniques. Although bulky multilens sets can correct these aberrations, they are impractical for wearable devices, limiting the potential for vivid near-eye displays. To address this challenge, researchers have explored the use of a single layer of metalens that can simultaneously correct colors and manipulate light~\cite{li2021meta}. However, the integration of such a system with geometrical optics remains unexplored, which is crucial for applications involving ray tracing in other parts of the imaging system.

Recent advancements have demonstrated that careful design of meta-atoms on a metalens enables precise control of the phase for discrete wavelengths~\cite{overvig2019dielectric}. Fortunately, in VR displays, where the light source can be restricted to several discrete bands (e.g., red, green, and blue), it becomes feasible to correct color aberrations and other lens-induced aberrations. This flexibility offers significant advantages at the system level, enhancing the overall performance of the VR experience.

To optimize the meta-atoms on the metalens, a differentiable Rigorous Coupled Wave Analysis (RCWA) simulator is required~\cite{zhu2020differentiable}. Although differentiable RCWA solvers are now widely available, their computational speed is still limited compared to wave-ray conversion and ray tracing methods. Moreover, since a metalens typically consists of thousands of meta-atoms, simulating each one at every iteration is computationally infeasible. To overcome these challenges, we initially train a surrogate model to characterize how the shape (represented by four parameters) and height of the meta-atoms affect the correction phase and wavefront for the aberrations. In this study, we focus on rectangular meta-atoms and simulate the effective indices of different colors (wavelengths of 488 nm, 532 nm, and 658 nm) from the fundamental mode. The effective phase modulation is calculated as $\phi_{\text{mod}} = k_0 n_{\text{eff}} h$.

\begin{figure*}[!htp]
    \centering
    \includegraphics[width=0.95\textwidth]{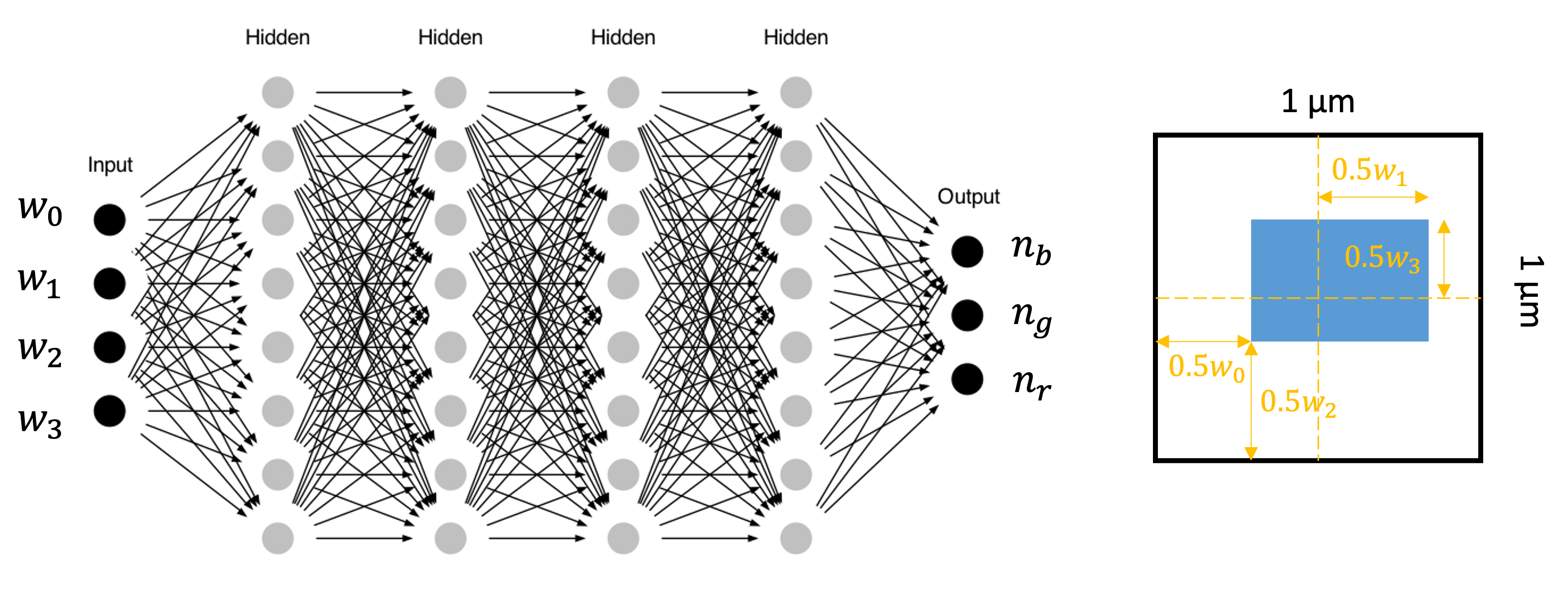}
    \caption{The architechture of the surrogate neural network and parameterization. We input four parameters, and use the simulated response for the three colors (blue, green, red) to train a simple multi-layer SIREN. To avoid overfitting, we choose a rather simple structure. The diagram of the cross-section of a meta-atam is shown in the right.}\label{fig:surrogate}
\end{figure*}

For the surrogate model, we employ a SIREN neural network architecture consisting of four hidden layers, each with eight hidden neurons. The input parameters are four meta-atom shape descriptors ($w_0$, $w_1$, $w_2$, and $w_3$), while the outputs are the effective indices ($n_b$, $n_g$, and $n_r$) corresponding to blue, green, and red light, respectively. The meta-atoms can be fabricated using silicon, surrounded by SiO2. To train the neural network, we simulate the results for 10,000 different configurations and subsequently evaluate its performance using an additional 1,000 randomly generated configurations. The loss is the summation of the squared distances between the outputs and the groundtruth effective indices of different colors. We train the neural network for 5,000 iterations. The final training loss is $7.5\times 10^{-4}$ while the testing loss is $9.8\times 10^{-4}$, which are accurate enough for our tasks. The architechture of the neural network and the shape of the meta-atoms are illustrated in \figref{surrogate}.

\begin{figure*}[!htp]
    \centering
    \includegraphics[width=0.9\textwidth]{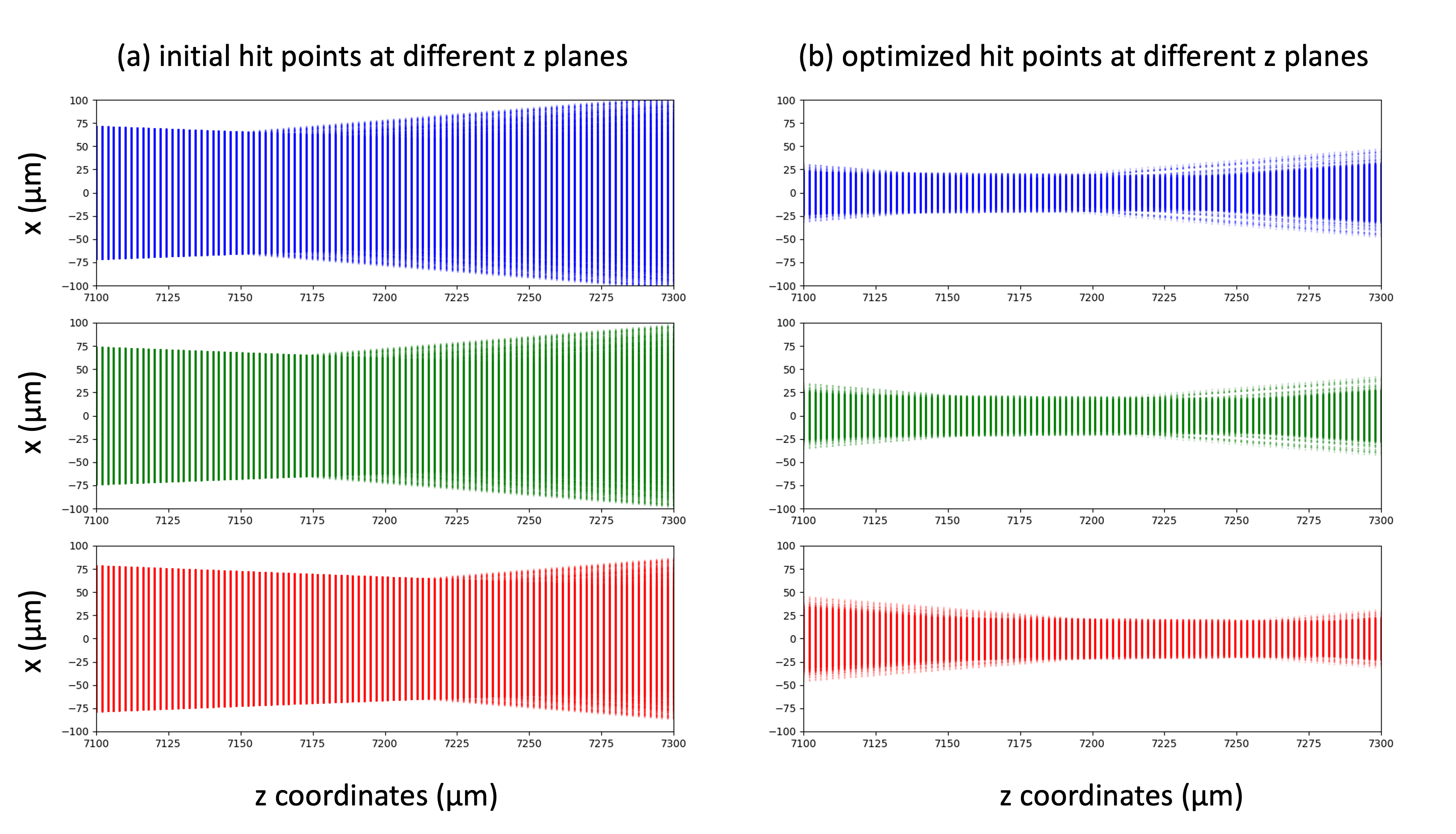}
    \caption{The initial and optimized hit points for different $z$-planes. Initially, rays of different colors focus at different positions, and the focus quality is limited because of geometrical aberrations. After optimization, rays of different colors all get better focus quality at $z = 7.2~{\rm mm}$. }\label{fig:surrogate_hitpoint}
\end{figure*}

\begin{figure*}[!htp]
    \centering
    \includegraphics[width=0.95\textwidth]{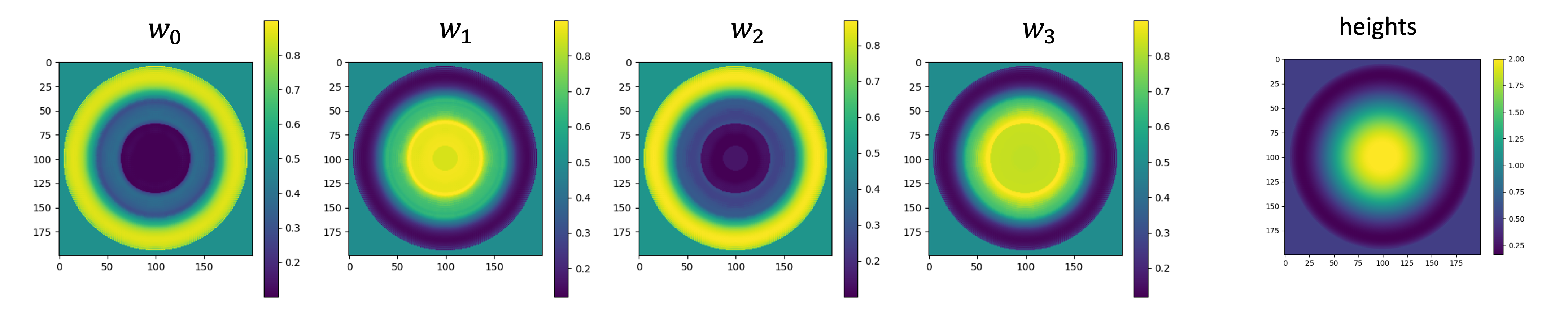}
    \caption{The optimized parameters and heights of the meta-atoms. The units in the figures are $\um$s. The size of metalens is $2\times 2~{\rm mm^2}$. Each meta-atom should have a period of $1~\um$, so we need to set the resolution of metalens to be $2000\times 2000$. Because of the limitation of GPU memory, we optimize for a lower resolution $200\times 200$. For fabrication, we can interpolate the results to get the values for a higher resolution. }\label{fig:meta_atom_optimized_parameters}
\end{figure*}

We start with a single-sided spherical lens with curvature radius $2.1~{\rm mm}$. The lens is located at $z = 2~{\rm mm}$. From Lensmaker's equation, the focal point should be at 
\begin{equation}
    f = z + \frac{R}{n-1} = 2 + \frac{2.1}{1.46 - 1} \approx 6.56 ~{\rm mm}
\end{equation}
However, when we simulate the system through ray tracing, because of lens aberrations, the actual focal point is near $7.2~{\rm mm}$. Furthermore, rays of different colors focus at different positions, as is shown in \figref{surrogate_hitpoint}(a). Based on that, we add the SIREN-surrogated metalens at $z = 0~{\rm mm}$ to correct the errors. After the wave-ray co-optimization, the final lens radius is $2098.88~\um$. And all colors get focused at $z = 7.2~{\rm mm}$ with minimum color aberrations, which is shown in \figref{surrogate_hitpoint}(b). The four optimized parameters and the height of meta-atoms are shown in \figref{meta_atom_optimized_parameters}

It has been demonstrated that by jointly optimizing the metalens and the geometrical lens, significant improvements can be achieved in terms of focus quality and color aberration reduction.
\section{Conclusion}
To summarize, our work proposes a fully differentiable framework for co-optimizing metalens and geometrical lens, which is the first attempt to combine ray tracing and wavefront modulation in a gradient-based optimization approach. However, our framework is based on certain assumptions, such as neglecting light diffraction and using the geometrical point spread function as the response of the optical system.

In future work, we plan to optimize the end-to-end framework over a large image dataset to connect neural networks for more domain-specific tasks. We also acknowledge recent advancements in path tracing techniques, such as~\cite{steinberg2021generic}, which can handle wave effects more effectively. It will be interesting to explore differentiable optimization approaches for such techniques for the purpose of optical design.

\section*{Acknowledgments} We thank Zhao Dong, and Shuang Zhao for their valuable suggestions.

\bibliography{main}





\end{document}